\documentstyle[12pt,amssymb,amsmath]{article}
\newtheorem{theorem}{Theorem}[section]
\newtheorem{lemma}{Lemma}[section]

\begin{document}

\title{ {\bf A new constrained mKP hierarchy and the generalized Darboux transformation for the mKP equation with self-consistent sources  }  }
\author{ {\bf  Ting Xiao
    \hspace{1cm} Yunbo Zeng\dag } \\
    {\small {\it
    Department of Mathematical Sciences, Tsinghua University,
    Beijing 100084, China}} \\
    {\small {\it \dag
     Email: yzeng@math.tsinghua.edu.cn}}}

\date{}
\maketitle
\renewcommand{\theequation}{\arabic{section}.\arabic{equation}}

\begin{abstract}
The mKP equation with self-consistent sources (mKPESCS) is treated
in the framework of the constrained mKP hierarchy. We introduce a
new constrained mKP hierarchy which may be viewed as the
stationary hierarchy of the mKP hierarchy with self-consistent
sources. This offers a natural way to obtain the Lax
representation for the mKPESCS. Based on the conjugate Lax pairs,
we construct the generalized Darboux transformation with arbitrary
functions in time $t$ for the mKPESCS which, in contrast with the
Darboux transformation for the mKP equation, provides a
non-auto-B\"{a}cklund transformation between two mKPESCSs with
different degrees. The formula for $n$-times repeated generalized
Darboux transformation is proposed and enables us to find the
rational solutions (including the lump solutions), soliton
solutions and the solutions of breather type of the mKPESCS.
\end{abstract}

\hskip\parindent


\section{Introduction}
\setcounter{equation}{0} \hskip\parindent Soliton equations with
self-consistent sources (SESCSs) are important models in many
fields of physics, such as hydrodynamics, solid state physics,
plasma physics, etc. [1-8,15]. For example, the nonlinear
Schr\"{o}dinger equation with self-consistent sources represents
the nonlinear interaction of an electrostatic high-frequency wave
with the ion acoustic wave in a two component homogeneous
plasma\cite{Claude91}. The KdV equation with self-consistent
sources describes the interaction of long and short
capillary-gravity waves\cite{Leon90(1)}. The KP equation with
self-consistent sources describes the interaction of a long wave
with a short-wave packet propagating on the x,y plane at an angle
to each other(see \cite{Mel'nikov87} and the references therein).
Until now, much development has been made in the study of SESCS.
For example, in (1+1)-dimensional case, some SESCSs such as the
KdV, modified KdV, nonlinear Schr\"{o}dinger, AKNS and Kaup-Newell
hierarchies with self-consistent sources were solved by the
inverse scattering method [1,2,3,6-10]. Also a type of generalized
binary Darboux transformations with arbitrary functions in time
$t$ for some (1+1)-dimensional SESCSs , which offer a
non-auto-B\"{a}cklund transformation between two SESCSs with
different degrees of sources, have been constructed and can be
used to obtain N-soliton, positon and negaton solution [12-14]. In
(2+1)-dimensional case, some results to the SESCSs have been
obtained. The soliton solution of the KP equation with
self-consistent sources (KPESCS) was first found by Mel'nikov
\cite{Mel'nikov87,Mel'nikov89(2)}. However, since the explicit
time part of the Lax representation of the KPESCS was not found,
the method to solve the KPESCS by inverse scattering
transformation in \cite{Mel'nikov87,Mel'nikov89(2)} was quite
complicated. In \cite{XiaoTing2004}, in the framework of the
constrained KP hierarchy, we get the Lax representation of the KP
equation with self-consistent sources naturally and construct the
generalized binary Darboux transformation for it naturally. The
KPESCS is also studied by Hirota method in \cite{Zhangdajun2003}.

In this paper, we develop the idea presented in
\cite{XiaoTing2004} to study the mKP equation with self-consistent
sources. First we give a new constrained mKP hierarchy which may
be viewed as the stationary hierarchy of the mKP hierarchy with
self-consistent sources. This gives a natural way to find the Lax
representation for the mKPESCS. Using the conjugate Lax pairs, we
construct the generalized Darboux transformation with arbitrary
functions in time $t$ for the mKPESCS. In contrast with the
Darboux transformation for the mKP equation which offers a
 B\"{a}cklund transformation, this transformation
provides a non-auto-B\"{a}cklund transformation between two
mKPESCSs with different degrees of sources. By this generalized
Darboux transformation, some interesting solutions of mKPESCS such
as soliton solutions, rational solutions (including lump
solutions) and solutions of breather type are obtained.

The paper will be organized as follows. We recall some facts about
the mKP hierarchy and mKP equation through the pseudo-differential
operator (PDO) formalism in the next section. In section 3,  we
introduce a new constrained mKP hierarchy and give some examples
of equations. In section 4, we reveal the relation between the mKP
hierarchy with self-consistent sources and the constrained mKP
hierarchy given in the previous section. Then the conjugate Lax
pairs of the mKP hierarchy with self-consistent sources can be
obtained naturally. Using the conjugate Lax pairs, we can
construct the generalized Darboux transformations with arbitrary
functions in time for the mKPESCS. In Section 5, the n-times
repeated generalized Darboux transformation will be constructed by
which some interesting solutions for the mKPESCS are obtained in
section 6.
\section{The mKP hierarchy and the mKP equation}
\setcounter{equation}{0} \hskip\parindent
Let us consider the following
pseudo-differential operator(PDO)
\begin{equation}
\label{11}
    L=L_{mKP}=\partial+V+V_1\partial^{-1}+V_2\partial^{-2}+...,
\end{equation}
where $\partial$ denotes $\frac{\partial}{\partial x}$, and
$V,V_j,j=1,...$ are functions. Denote $B_m=(L^m)_{ \geq1}$ for
$\forall m \in N$ where $(L^m)_{ \geq1}$ represents the projection
of $L^m$ to its differential part whose order is more than 1. Then
the mKP hierarchy is defined as \cite{Oevel93}
\begin{equation}
\label{12}
    L_{t_k}=[B_k,L], k \geq 1.
\end{equation}
or the equivalent form
\begin{equation}
\label{13}
    (L^n)_{t_k}=[B_k,L^n], n, k \geq 1.
\end{equation}
The mKP hierarchy (\ref{12}) can also be written in the zero-curvature form
\begin{equation}
\label{14}
    (B_n)_{t_m}-(B_m)_{t_n}+[B_n,B_m] = 0,\ \ n,m\geq2.
\end{equation}
The equation (\ref{14}) has a pair of conjugate Lax pairs as
follows
\begin{subequations}
\label{15}
\begin{equation}
\label{15.1}
    \psi_{1,t_m} = (B_m\psi_1),
\end{equation}
\begin{equation}
\label{15.2}
   \psi_{1,t_n} = (B_n\psi_1),
\end{equation}
\end{subequations}
and
\begin{subequations}
\label{16}
\begin{equation}
\label{16.1}
    \psi_{2,t_m} = (\tilde{B}_m\psi_2),
\end{equation}
\begin{equation}
\label{16.2}
    \psi_{2,t_n} = (\tilde{B}_n\psi_2),
\end{equation}
\end{subequations}
where $\tilde{B_k}=-(\partial B_k\partial^{-1})^*$, $k \geq2$. We
make a convention that for any operator $P$ and a function $f$,
$(Pf)$ means that the operator $P$ acts on $f$ while $Pf$ means
the product of $P$ and $f$. It is easy to see that $\tilde{B_k}$
are also differential operators. When $n=2,\ m=3$ we get the mKP
equation as follows
\begin{equation}
\label{17}
    4V_{t_3}-V_{xxx}+6V^2V_x-3(D^{-1}V_{t_2t_2})-6V_x(D^{-1}V_{t_2})=0
\end{equation}
where $DD^{-1}=D^{-1}D=1$. Set
\begin{equation}
\label{18}
    u=-V,\ \  t=-\frac{1}{4}t_3,\ \  y=\alpha t_2,
\end{equation}
the mKP equation will be written as
\begin{equation}
\label{19}
    u_t-6u^2u_x+u_{xxx}+3\alpha^2(D^{-1}u_{yy})-6\alpha u_x(D^{-1}u_y) =0,
\end{equation}
which is called the mKPI equation when $\alpha=i$ and mKPII equation when
$\alpha=1$. From (\ref{15}) and (\ref{16}), we will get the conjugate Lax
pairs of (\ref{19}) respectively as follows
\begin{subequations}
\label{110}
\begin{equation}
\label{110.1}
    \alpha\psi_{1,y}=\psi_{1,xx}-2u\psi_{1,x},
\end{equation}
\begin{equation}
\label{110.2}
    \psi_{1,t}=(A_1(u)\psi_1),\ \ \
    A_1(u)=-4\partial^3+12u\partial^2-6(-u_x+u^2-\alpha D^{-1}u_y)\partial,
\end{equation}
\end{subequations}
and
\begin{subequations}
\label{111}
\begin{equation}
\label{111.1}
    \alpha\psi_{2,y}=-\psi_{2,xx}-2u\psi_{2,x},
\end{equation}
\begin{equation}
\label{111.2}
     \psi_{2,t}=(A_2(u)\psi_2),\ \ \
    A_2(u)=-4\partial^3-12u\partial^2-6(u_x+u^2-\alpha D^{-1}u_y)\partial,
\end{equation}
\end{subequations}

It is known that the system (\ref{110}) is covariant w.r.t. the
following transformations \cite{Estevez}
\begin{subequations}
\label{112}
\begin{equation}
\label{112.1}
    \psi_1[1]=\psi_1-f_1\frac{\int\psi_{1,x}g_1{\mathrm{d}}x+C_2}{\int
f_{1,x}g_1{\mathrm{d}}x-C_1},
\end{equation}
\begin{equation}
\label{112.2}
    u[1]=u+\partial_x {\mathrm{ln}} \frac{\int g_{1,x}f_1{\mathrm{d}}x+C_1}{\int
f_{1,x}g_1{\mathrm{d}}x-C_1},
\end{equation}
\end{subequations}
while the system (\ref{111}) is covariant w.r.t.
\begin{subequations}
\label{113}
\begin{equation}
\label{113.1}
    \psi_2[1]=\psi_2-g_1\frac{\int\psi_{2,x}f_1{\mathrm{d}}x+C_2}{\int
g_{1,x}f_1{\mathrm{d}}x+C_1},
\end{equation}
\begin{equation}
\label{113.2}
    u[1]=u+\partial_x{\mathrm{ln}}\frac{\int g_{1,x}f_1{\mathrm{d}}x+C_1}{\int f_{1,x}g_1{\mathrm{d}}x-C_1},
\end{equation}
\end{subequations}
where $f_1,g_1$ are solutions of (\ref{110}) and (\ref{111})
respectively and $C_1, C_2$ are arbitrary constants. We point out
that throughout the paper, the integral operation $\int f_1f_2
{\mathrm{d}}x$ means $\int_{-\infty}^x f_1f_2 {\mathrm{d}}x$ or
$-\int_x^{\infty} f_1f_2 {\mathrm{d}}x$ and contains no arbitrary
function of $y$ and $t$, only numerical constant if we impose some
suitable boundary condition on the integrand functions $f_1$ and
$f_2$ at $x=-\infty$ or $x=\infty$. Substituting (\ref{112.1})
(with $C_2=0$,$C_1=C$) into (\ref{110.2}), we will get the
following identity
\begin{equation}
\label{114}
\begin{array}{ll}
   &(A_1(u[1])\psi_1[1])\\
=&(\psi_1-f_1\frac{\int\psi_{1,x}g_1{\mathrm{d}}x}{\int
f_{1,x}g_1{\mathrm{d}}x-C})_t\\
=&\psi_{1,t}-f_{1,t}\frac{\int\psi_{1,x}g_1{\mathrm{d}}x}{\int
f_{1,x}g_1{\mathrm{d}}x-C}-f_1\frac{\int(\psi_{1,x}g_{1,t}+\psi_{1,xt}g_1){\mathrm{d}}x(\int
f_{1,x}g_1{\mathrm{d}}x-C)-(\int\psi_{1,x}g_1{\mathrm{d}}x)[\int(g_{1,t}f_{1,x}+g_1f_{1,xt}){\mathrm{d}}x]}{(\int
f_{1,x}g_1{\mathrm{d}}x-C)^2}\\
=&(A_1(u)\psi_1)-(A_1(u)f_1)\frac{\int\psi_{1,x}g_1{\mathrm{d}}x}{\int
f_{1,x}g_1{\mathrm{d}}x-C}\\
&-f_1\frac{[\int(\psi_{1,x}(A_2(u)g_1)+(A_1(u)\psi_1)_xg_1){\mathrm{d}}x](\int
f_{1,x}g_1{\mathrm{d}}x-C)-(\int
\psi_{1,x}g_1{\mathrm{d}}x)[\int((A_2(u)g_1)f_{1,x}+g_1(A_1(u)f_1)_x){\mathrm{d}}x]}{(\int
f_{1,x}g_1{\mathrm{d}}x-C)^2}.
\end{array}
\end{equation}

\section{A new constraint of the mKP hierarchy}
\setcounter{equation}{0} \hskip\parindent In \cite{Oevel93},
W.Oevel and W.Strampp have studied the constraint of the PDO $L$
(\ref{11}) as
\begin{equation}
\label{20}
     L^n=(L^n)_{\geq 1}+v_0+\partial^{-1}\psi,
\end{equation}
from which we will get the Kaup-Broer hierarchy when $n=1$. Here
we consider a new constraint as follows
\begin{equation}
\label{21}
     L^n=(L^n)_{\geq 1}+q\partial^{-1}r\partial.
\end{equation}
where $q$,$r$ satisfy that
\begin{equation}
\label{22}
  q_{t_k}=(B_kq),\ \ \  r_{t_k}=(\tilde{B}_kr),
\end{equation}
and $B_k=((L^n)^{\frac{k}{n}})_{\geq 1}=[((L^n)_{\geq
1}+q\partial^{-1}r\partial)^{\frac{k}{n}}]_{\geq 1}$.\\
Then a new $n$-constrained mKP hierarchy will be obtained as
\begin{subequations}
\label{23}
\begin{equation}
\label{23.1}
    (L^n)_{t_k}=[(L^k)_{\geq 1},L^n]=[B_k,L^n],
\end{equation}
\begin{equation}
\label{23.2}
    q_{t_k}=(B_kq),
\end{equation}
\begin{equation}
\label{23.3}
    r_{t_k}=(\tilde{B}_kr),
\end{equation}
\end{subequations}
First, we will prove that the constraint (\ref{21}) together with
the condition (\ref{22}) is compatible with the mKP hierarchy
(\ref{12}). The following formulas for PDO will be useful in the
proof and we list them below,
\begin{subequations}
\label{24}
\begin{equation}
\label{24.1} (\Lambda^*)_0=res(\partial^{-1}\Lambda),\ \
(\Lambda)_0=res(\Lambda\partial^{-1}),\ \
(\Lambda\partial^{-1})_{<0}=(\Lambda)_0\partial^{-1}+(\Lambda)_{<0}\partial^{-1},
\end{equation}
\begin{equation}
\label{24.2}
    (Pq\partial^{-1}r)_{<0}=(Pq)\partial^{-1}r,\ \ \ (q\partial^{-1}rP)_{<0}=q\partial^{-1}(P^*r),
\end{equation}
\end{subequations}
where $\Lambda$ is an arbitrary PDO and P differential operator.
$(A)_0$ denote the zero order term for a PDO $A$.
\begin{theorem}
The constraint (\ref{21}) together with the condition (\ref{22})
is compatible with the mKP hierarchy (\ref{12}).
\end{theorem}
Proof: We need to prove the following identity
\begin{equation}
\label{25}
    (q\partial^{-1}r\partial)_{t_k}=[B_k,L^n]_{\leq 0}=[B_k,q\partial^{-1}r\partial]_{\leq 0},
\end{equation}
\begin{equation}
\label{26}
\begin{array}{rcl}
\text{the l.h.s. of}
(\ref{25})&=&q_{t_k}\partial^{-1}r\partial+q\partial^{-1}r_{t_k}\partial\\
&=&(B_kq)\partial^{-1}r\partial+q\partial^{-1}(\tilde{B}_kr)\partial\\
&\stackrel{\triangle}=&l_1+l_2\\
\end{array}
\end{equation}
\begin{equation}
\label{27}
\begin{array}{rcl}
\text{the r.h.s. of}
(\ref{25})&=&(B_k q\partial^{-1}r\partial)_{\leq0}-(q\partial^{-1}r\partial B_k)_{\leq0}\\
&\stackrel{\triangle}=&r_1-r_2\\
\end{array}
\end{equation}
\begin{equation}
(l_1)_0=((B_kq)\partial^{-1}r\partial)_0=(B_kq)r,\ \ \
(l_2)_0=(q\partial^{-1}(\tilde{B}_kr)\partial)_0=q(\tilde{B}_kr),
\end{equation}
\begin{equation}
\begin{array}{rcl}
(r_1)_0&=&res[\partial^{-1}(r_1)^*]=res[\partial^{-1}(\partial
r\partial^{-1} qB_k^*)]=res(r\partial^{-1}
qB_k^*)\\
&=&res(r\partial^{-1} qB_k^*)_{<0}=res(r\partial^{-1}
(B_kq))=r(B_kq),
\end{array}
\end{equation}
\begin{equation}
\begin{array}{rcl}
(r_2)_0&=&(q\partial^{-1}r\partial
B_k)_0=res[q\partial^{-1}r\partial
B_k\partial^{-1}]=res[q\partial^{-1}r(\partial
B_k\partial^{-1})]\\
&=&res[q\partial^{-1}((\partial B_k\partial^{-1})^*r)]=q((\partial
B_k\partial^{-1})^*r)=-q(\tilde{B}_kr),
\end{array}
\end{equation}
So
\begin{equation}
\label{28}
(l_1)_0+(l_2)_0=(r_1)_0-(r_2)_0.
\end{equation}
\begin{equation}
(l_1)_{<0}=((B_kq)\partial^{-1}r\partial)_{<0}=-(B_kq)\partial^{-1}r_x,
\end{equation}
\begin{equation}
(l_2)_{<0}=(q\partial^{-1}(\tilde{B}_kr)\partial)_{<0}=-q\partial^{-1}[\partial((\tilde{B}_kr))]=q\partial^{-1}[\partial\partial^{-1}(B_k^*\partial
r)]=q\partial^{-1}(B_k^*r_x),
\end{equation}
By the last formula of (\ref{24.1}), we have
\begin{equation}
(r_1\partial^{-1})_{<0}=(B_kq\partial^{-1}r)_{<0}=(r_1)_0\partial^{-1}+(r_1)_{<0}\partial^{-1},
\end{equation}
i.e.
$$(B_k q)\partial^{-1}r=(r_1)_0\partial^{-1}+(r_1)_{<0}\partial^{-1}.$$
Multiplying $\partial$ on the right and taking the negative part
of both sides of the above identity, we get
$$((B_kq)\partial^{-1}r\partial)_{<0}=(r_1)_{<0}$$
So
$$(r_1)_{<0}=-(B_kq)\partial^{-1}r_x.$$
\begin{equation}
(r_2)_{<0}=(q\partial^{-1}r\partial
B_k)_{<0}=q\partial^{-1}(\partial
B_k)^*(r)=-q\partial^{-1}(B_k^*r_x)
\end{equation}
So we have
\begin{equation}
\label{29} (l_1)_{<0}+(l_2)_{<0}=(r_1)_{<0}-(r_2)_{<0}
\end{equation}
From (\ref{28}) and (\ref{29}), we can see (\ref{25}) holds.\\
This completes the proof.\\

We give some examples below.\\
(a) \ $1$-constraint($n=1$).\\
Here
\begin{equation}
     L=\partial+q\partial^{-1}r\partial.
\end{equation}
So
\begin{equation}
     V=qr,\ \ V_1=-qr_x,\ \ ...
\end{equation}
$$B_2=(L^2)_{\geq 1}=\partial^2+2qr\partial,\ \ \ \ \ \tilde{B}_2=-(\partial B_2\partial^{-1})^*=-\partial^2+2qr\partial,$$
$$B_3=(L^3)_{\geq 1}=\partial^3+3qr\partial^2+(3q^2r^2+3q_xr)\partial,$$
$$\tilde{B}_3=-(\partial B_3\partial^{-1})^*=\partial^3-3qr\partial^2+(3q^2r^2-3qr_x)\partial,\ \ ...$$
The first two equations of the $1$-constrained hierarchy are
\begin{subequations}
\label{210}
\begin{equation}
\label{210.1} q_{t_2}=q_{xx}+2qrq_x,
\end{equation}
\begin{equation}
\label{210.2} r_{t_2}=-r_{xx}+2qrr_x,
\end{equation}
\end{subequations}
and
\begin{subequations}
\label{211}
\begin{equation}
\label{211.1} q_{t_3}=q_{xxx}+3qrq_{xx}+(3q^2r^2+3q_xr)q_x,
\end{equation}
\begin{equation}
\label{211.2} r_{t_3}=r_{xxx}-3qrr_{xx}+(3q^2r^2-3qr_x)r_x.
\end{equation}
\end{subequations}
Equation (\ref{210}) is the generalized NS equation with
derivative coupling given by Chen et al \cite{Chen79,ChengYi92}.
The
constrained hierarchy is also studied in \cite {Gengxianguo99}.\\
(b) \ $2$-constraint($n=2$).\\
Here
\begin{equation}
     L^2=\partial^2+2V\partial+q\partial^{-1}r\partial.
\end{equation}
from which we find
$$V_1=qr-V_x-V^2,\ \ \ \ \ \ ...$$
$$B_2=(L^2)_{\geq 1}=\partial^2+2V\partial,\ \ \ \ \ \tilde{B}_2=-(\partial B_2\partial^{-1})^*=-\partial^2+2V\partial,$$
$$B_3=(L^3)_{\geq 1}=\partial^3+3V\partial^2+3qr\partial,\ \ \ \ \ \tilde{B}_3=-(\partial B_3\partial^{-1})^*=\partial^3-3V\partial^2+(3qr-3V_x)\partial,\ \ ...$$
The first two equations of the $2$-constrained hierarchy are
\begin{subequations}
\label{212}
\begin{equation}
\label{212.1} V_{t_2}=(qr)_x,
\end{equation}
\begin{equation}
\label{212.2} q_{t_2}=q_{xx}+2Vq_x,
\end{equation}
\begin{equation}
\label{212.3} r_{t_2}=-r_{xx}+2Vr_x.
\end{equation}
\end{subequations}
and
\begin{subequations}
\label{213}
\begin{equation}
\label{213.1}
V_{t_3}=V_{xxx}+3VV_{xx}+6V^2V_x+3qrV_x-\frac{3}{2}(qr_x)_x,
\end{equation}
\begin{equation}
\label{213.2} q_{t_3}=q_{xxx}+3Vq_{xx}+3qrq_x,
\end{equation}
\begin{equation}
\label{213.3} r_{t_3}=r_{xxx}-3Vr_{xx}+(3qr-3V_x)r_x.
\end{equation}
\end{subequations}
(c) \ $3$-constraint($n=3$).\\
Here
\begin{equation}
     L^3=\partial^3+3V\partial^2+3(V^2+V_x+V_1)\partial+q\partial^{-1}r\partial.
\end{equation}
The first equation of the $3$-constrained hierarchy is
\begin{subequations}
\label{214}
\begin{equation}
\label{214.1} V_{t_2}=V_{xx}+2V_{1,x}+2VV_x,
\end{equation}
\begin{equation}
\label{214.2}
3V_{1,t_2}=-2V_{xxx}-6VV_{xx}-6V^2V_x-6V_x^2-3V_{1,xx}-6VV_{1,x}-6V_1V_x+2(qr)_x,
\end{equation}
\begin{equation}
\label{214.3} q_{t_2}=q_{xx}+2Vq_x,
\end{equation}
\begin{equation}
\label{214.4} r_{t_2}=-r_{xx}+2Vr_x.
\end{equation}
\end{subequations}
Eliminating $V_1$ from the above equation, we get
\begin{subequations}
\label{215}
\begin{equation}
\label{215.1}
\frac{1}{2}V_{xxx}+\frac{3}{2}D^{-1}(V_{yy})+3(D^{-1}V_y)V_x-3V^2V_x-2(qr)_x=0,
\end{equation}
\begin{equation}
\label{215.2}
 q_{t_2}=q_{xx}+2Vq_x,
\end{equation}
\begin{equation}
\label{215.3} r_{t_2}=-r_{xx}+2Vr_x.
\end{equation}
\end{subequations}

\section{The mKP equation with self-consistent sources and its generalized Darboux transformation}
\setcounter{equation}{0} \hskip\parindent If generalizing the
constraint (\ref{21}) to
\begin{equation}
\label{31}
     L^n=(L^n)_{\geq 1}+\sum_{i=1}^{N}q_i\partial^{-1}r_i\partial.
\end{equation}
where
\begin{equation}
\label{32}
   q_{i,t_k}=(B_kq_i),\ \ \  r_{i,t_k}=(\tilde{B}_kr_i),
\end{equation}
and adding the term $(B_k)_{t_n}$ to the right hand side of
(\ref{23.1}), we can define the mKP hierarchy with self-consistent
sources as follows
\begin{subequations}
\label{33}
\begin{equation}
\label{33.1}
    (B_k)_{t_n}-(L^n)_{t_k}+[B_k,L^n]=0,
\end{equation}
\begin{equation}
\label{33.2}
    q_{i,t_k}=(B_kq_i),
\end{equation}
\begin{equation}
\label{33.3}
    r_{i,t_k}=(\tilde{B}_kr_i).
\end{equation}
\end{subequations}
So if the variable "$t_n$" is viewed as the evolution variable,
the $n$-constrained mKP hierarchy may be regarded as the
stationary hierarchy of the mKP hierarchy with self-consistent
sources. Under the condition (\ref{33.2}) and (\ref{33.3}), we
naturally get the conjugate Lax pairs of (\ref{33.1}) as follows
\begin{subequations}
\label{34}
\begin{equation}
\label{34.1}
    \psi_{1,t_k}=(B_k\psi_1),
\end{equation}
\begin{equation}
\label{34.2}
    \psi_{1,t_n}=(L^n\psi_1)=(B_n\psi_1)+\sum_{i=1}^{N}q_i\int r_i\psi_{1,x}{\mathrm{d}}x,
\end{equation}
\end{subequations}
and
\begin{subequations}
\label{35}
\begin{equation}
\label{35.1}
    \psi_{2,t_k}=(\tilde{B}_k\psi_2),
\end{equation}
\begin{equation}
\label{35.2}
    \psi_{2,t_n}=(\tilde{L}^n\psi_2)=(\tilde{B}_n\psi_2)-([\partial(\sum_{i=1}^{N}q_i\partial^{-1}r_i\partial)\partial^{-1}]^*\psi_2)=(\tilde{B}_n\psi_2)-\sum_{i=1}^{N}r_i\int q_i\psi_{2,x}{\mathrm{d}}x,
\end{equation}
\end{subequations}
When $n=3$, $k=2$, under the transformation (\ref{18}) and setting
$$\Phi_i=r_i,\ \ \ \ \ \ \Psi_i=q_i,$$
we will get the mKP equation with self-consistent sources
(mKPESCS) and its conjugate Lax pairs respectively from
(\ref{33}),(\ref{34}) and (\ref{35}).\\
The mKPESCS is
\begin{subequations}
\label{36}
\begin{equation}
\label{36.1}
    u_t+u_{xxx}+3\alpha^2D^{-1}(u_{yy})-6\alpha D^{-1}(u_y)u_x-6u^2u_x+4\sum_{i=1}^{N}(\Psi_i\Phi_i)_x=0,
\end{equation}
\begin{equation}
\label{36.2}
    \alpha\Psi_{i,y}=\Psi_{i,xx}-2u\Psi_{i,x},
\end{equation}
\begin{equation}
\label{36.3}
    \alpha\Phi_{i,y}=-\Phi_{i,xx}-2u\Phi_{i,x},
\end{equation}
\end{subequations}
 which is called the mKPIESCS when $\alpha=i$ and mKPIIESCS when $\alpha=1$. Under the condition (\ref{36.2}) and (\ref{36.3}),
the conjugate Lax pairs for (\ref{36.1}) are
\begin{subequations}
\label{37}
\begin{equation}
\label{37.1}
    \alpha\psi_{1,y}=\psi_{1,xx}-2u\psi_{1,x},
\end{equation}
\begin{equation}
\label{37.2}
    \psi_{1,t}=(A_1(u)\psi_1)+T_N^1(\Psi,\Phi)\psi_1,\ \
    T_N^1(\Psi,\Phi)\psi_1=-4\sum_{i=1}^{N}\Psi_i\int
    \Phi_i\psi_{1,x}{\mathrm{d}}x,
\end{equation}
\end{subequations}
and
\begin{subequations}
\label{38}
\begin{equation}
\label{38.1}
    \alpha\psi_{2,y}=-\psi_{2,xx}-2u\psi_{2,x},
\end{equation}
\begin{equation}
\label{38.2}
    \psi_{2,t}=(A_2(u)\psi_2)+T_N^2(\Psi,\Phi)\psi_2,\ \
    T_N^2(\Psi,\Phi)\psi_2=4\sum_{i=1}^{N}\Phi_i\int
    \Psi_i\psi_{2,x}{\mathrm{d}}x,
\end{equation}
\end{subequations}
For the system (\ref{37}), we can construct the following Darboux
transformation.
\begin{theorem}
 Assume $u, \Phi_1,...,\Phi_N,\Psi_1,...,\Psi_N$ be a solution of the mKPESCS (\ref{36})
  and $f_1$, $g_1$ satisfy (\ref{37}) and (\ref{38}) respectively, then the
  system (\ref{37}) has the following Darboux transformation
\begin{subequations}
\label{39}
\begin{equation}
\label{39.1}
\psi_1[1]=\psi_1-f_1\frac{\int\psi_{1,x}g_1{\mathrm{d}}x}{\int
f_{1,x}g_1{\mathrm{d}}x-C},
\end{equation}
\begin{equation}
\label{39.2} u[1]=u+\partial_x {\mathrm{ln}}\frac{\int
g_{1,x}f_1{\mathrm{d}}x+C}{\int
f_{1,x}g_1{\mathrm{d}}x-C}=u+\frac{g_{1,x}f_1}{\int
g_{1,x}f_1{\mathrm{d}}x+C}-\frac{f_{1,x}g_1}{\int
f_{1,x}g_1{\mathrm{d}}x-C},
\end{equation}
\begin{equation}
\label{39.3}
 \Psi_j[1]=\Psi_j-f_1\frac{\int\Psi_{i,x}g_1{\mathrm{d}}x}{\int f_{1,x}g_1{\mathrm{d}}x-C},
\end{equation}
\begin{equation}
\label{39.4}
 \Phi_j[1]=\Phi_j-g_1\frac{\int f_1\Phi_{j,x}{\mathrm{d}}x}{\int f_1g_{1,x}{\mathrm{d}}x+C},\qquad
 j=1,...,N.
\end{equation}
\end{subequations}
where $C$ is an arbitrary constant.
\end{theorem}
{\bf{Proof}}: It is obvious that $u[1]$, $\psi_1[1]$,
$\Phi_i[1]$,$\Psi_i[1]$, $i=1,...,N$ satisfy
(\ref{36.2}),(\ref{36.3}) and (\ref{37.1}). So we only need to
prove that (\ref{37.2}) holds, i.e., to prove the following
equality
\begin{equation}
\label{310}
\psi_1[1]_t=(A_1(u[1])\psi_1[1])+T^1_N(\Psi[1],\Phi[1])\psi_1[1].
\end{equation}
Using (\ref{37.2}), we have
\begin{equation}
\begin{array}{lll}
\psi_1[1]_t
&=&(A_1(u)\psi_1)+T^1_N(\Psi,\Phi)\psi_1-((A_1(u)f_1)+T^1_N(\Psi,\Phi)f_1)\frac{\int\psi_{1,x}g_1{\mathrm{d}}x}{\int
f_{1,x}g_1{\mathrm{d}}x-C}
\\
&&-f_1\frac{\int((A_2(u)g_1)+T^2_N(\Psi,\Phi)g_1)\psi_{1,x}{\mathrm{d}}x+\int
g_1((A_1(u)\psi_1)+T^1_N(\Psi,\Phi)\psi_1)_x{\mathrm{d}}x}{\int
f_{1,x}g_1{\mathrm{d}}x-C}\\
&&+f_1(\int
g_1\psi_{1,x}{\mathrm{d}}x)\frac{\int((A_2(u)g_1)+T^2_N(\Psi,\Phi)g_1)f_{1,x}{\mathrm{d}}x+\int
g_1((A_1(u)f_1)+T^1_N(\Psi,\Phi)f_1)_x{\mathrm{d}}x}{(\int
f_{1,x}g_1{\mathrm{d}}x-C)^2}
\end{array}
\end{equation}
It is easy to verify that (\ref{114}) still holds now. So we only
need to prove the following identity
\begin{equation}
\label{311}
\begin{array}{ll}
&T^1_N(\Psi[1],\Phi[1])\psi_1[1]\\
=&T^1_N(\Psi,\Phi)\psi_1-T^1_N(\Psi,\Phi)f_1\frac{\int\psi_{1,x}g_1{\mathrm{d}}x}{\int
f_{1,x}g_1{\mathrm{d}}x-C}-f_1\frac{\int
T^2_N(\Psi,\Phi)g_1\psi_{1,x}{\mathrm{d}}x+\int
g_1(T^1_N(\Psi,\Phi)\psi_1)_x{\mathrm{d}}x}{\int
f_{1,x}g_1{\mathrm{d}}x-C}
\\
&+f_1(\int g_1\psi_{1,x}{\mathrm{d}}x)\frac{\int
T^2_N(\Psi,\Phi)g_1f_{1,x}{\mathrm{d}}x+\int
g_1(T^1_N(\Psi,\Phi)f_1)_x{\mathrm{d}}x}{(\int
f_{1,x}g_1{\mathrm{d}}x-C)^2}
\end{array}
\end{equation}
By substituting the expression of $T^1_N$ and $T^2_N$ in
(\ref{37.2}) and (\ref{38.2}), we find
\begin{equation}
\begin{array}{ll}
& \text{the r.h.s. of (\ref{311})}\\
=&-4\sum_{j=1}^N\Psi_j\int\Phi_j\psi_{1,x}{\mathrm{d}}x+4f_1\sum_{j=1}^N\frac{(\int
g_1\Psi_{j,x}{\mathrm{d}}x)(\int\Phi_j\psi_{1,x}{\mathrm{d}}x)}{\int
f_{1,x}g_1{\mathrm{d}}x-C}\\
&+4\sum_{j=1}^N(\Psi_j-f_1\frac{\int
g_1\Psi_{j,x}{\mathrm{d}}x}{\int
f_{1,x}g_1{\mathrm{d}}x-C})\frac{(\int
g_1\psi_{1,x}{\mathrm{d}}x)(\int\Phi_jf_{1,x}{\mathrm{d}}x)}{\int
f_{1,x}g_1{\mathrm{d}}x-C}.
\end{array}
\end{equation}
Then using (\ref{39}) and (\ref{37.2}), we can show that
\begin{equation}
\begin{array}{ll}
& \text{the l.h.s. of (\ref{311})}\\
=&-4\sum_{j=1}^N\Psi_j\int\Phi_j\psi_{1,x}{\mathrm{d}}x+4f_1\sum_{j=1}^N\frac{(\int
g_1\Psi_{j,x}{\mathrm{d}}x)(\int\Phi_j\psi_{1,x}{\mathrm{d}}x)}{\int
f_{1,x}g_1{\mathrm{d}}x-C}\\
&+4\sum_{j=1}^N(\Psi_j-f_1\frac{\int
g_1\Psi_{j,x}{\mathrm{d}}x}{\int
f_{1,x}g_1{\mathrm{d}}x-C})\frac{(\int
g_1\psi_{1,x}{\mathrm{d}}x)(\int\Phi_jf_{1,x}{\mathrm{d}}x)}{\int
f_{1,x}g_1{\mathrm{d}}x-C}\\
=&\text{the r.h.s. of (\ref{311})}.
\end{array}
\end{equation}
This completes the proof.\\

If $C$ is replaced by $C(t)$, an arbitrary function in time $t$ in
(\ref{39}), then (\ref{36.2}),(\ref{36.3}) and (\ref{37.1}) are
also covariant w.r.t. (\ref{39}), but (\ref{37.2}) is not
covariant w.r.t. (\ref{39}) any longer. In fact, we have the
following theorem.
\begin{theorem}
 Given $u$,$\Psi_1,...,\Psi_N,\Phi_1,...,\Phi_N$ a solution of
the mKPESCS (\ref{36}) and let $f_1$ and $g_1$ be solutions of the
system  (\ref{37}) and (\ref{38}) respectively, then the
transformation with $C(t)$ (an arbitrary function in $t$) defined
by
\begin{subequations}
\label{312}
\begin{equation}
\label{312.1}
\psi_1[1]=\psi_1-f_1\frac{\int\psi_{1,x}g_1{\mathrm{d}}x}{\int
f_{1,x}g_1{\mathrm{d}}x-C(t)},
\end{equation}
\begin{equation}
\label{312.2} u[1]=u+\partial_x {\mathrm{ln}}\frac{\int
g_{1,x}f_1{\mathrm{d}}x+C(t)}{\int
f_{1,x}g_1{\mathrm{d}}x-C(t)}=u+\frac{g_{1,x}f_1}{\int
g_{1,x}f_1{\mathrm{d}}x+C(t)}-\frac{f_{1,x}g_1}{\int
f_{1,x}g_1{\mathrm{d}}x-C(t)},
\end{equation}
\begin{equation}
\label{312.3}
 \Psi_j[1]=\Psi_j-f_1\frac{\int\Psi_{i,x}g_1{\mathrm{d}}x}{\int f_{1,x}g_1{\mathrm{d}}x-C(t)},
\end{equation}
\begin{equation}
\label{312.4}
 \Phi_j[1]=\Phi_j-g_1\frac{\int f_1\Phi_{j,x}{\mathrm{d}}x}{\int f_1g_{1,x}{\mathrm{d}}x+C(t)},\qquad
 j=1,...,N,
\end{equation}
\text{and}
\begin{equation}
\label{312.5}
 \Psi_{N+1}[1]=-\frac{1}{2}\frac{\sqrt{\dot{C}(t)}f_1}{\int
f_{1,x}g_1{\mathrm{d}}x-C(t)},\ \ \
\Phi_{N+1}[1]=\frac{1}{2}\frac{\sqrt{\dot{C}(t)}g_1}{\int
g_{1,x}f_1{\mathrm{d}}x+C(t)},
\end{equation}
\end{subequations}
transforms (\ref{36.2}),(\ref{36.3}) and (\ref{37}) respectively
into
\begin{subequations}
\label{313}
\begin{equation}
\label{313.1}
    \alpha\Psi_i[1]_y=\Psi_i[1]_{xx}-2u[1]\Psi_i[1]_x,
\end{equation}
\begin{equation}
\label{313.2}
    \alpha\Phi_i[1]_y=-\Phi_i[1]_{xx}-2u[1]\Phi_i[1]_x, \ \ \
    i=1,...,N+1,
\end{equation}
\begin{equation}
\label{313.3}
    \alpha\psi_1[1]_y=\psi_1[1]_{xx}-2u[1]\psi_1[1]_x,
\end{equation}
\begin{equation}
\label{313.4}
    \psi_1[1]_t=A_1(u[1])\psi_1[1]+T_{N+1}^1(\Psi[1],\Phi[1])\psi_1[1].\ \
\end{equation}
\end{subequations}
So $u[1]$, $\Psi_i[1]$, $\Phi_i[1]$, $i=1,...,N+1$ is a new
solution of the mKPESCS (\ref{36}) with degree $N+1$.
\end{theorem}
{\bf{Proof}}:\ Equations (\ref{313.1}), (\ref{313.2}) and
(\ref{313.3}) hold obviously. We only need to prove (\ref{313.4}).
Substituting (\ref{312.1}) into the left hand side of
(\ref{313.4}) and using the result of the previous theorem, we
have
\begin{equation}
\psi_1[1]_t=(\psi_1-f_1\frac{\int \psi_{1,x}g_1{\mathrm{d}}x}{\int
f_{1,x}g_1{\mathrm{d}}x-C(t)})_t=A_1(u[1])\psi_1[1]+T_{N}^1(\Psi[1],\Phi[1])\psi_1[1]-\frac{\dot{C}(t)f_1\int
\psi_{1,x}g_1{\mathrm{d}}x }{(\int
f_{1,x}g_1{\mathrm{d}}x-C(t))^2},
\end{equation}
So we only need to prove
$$-4\Psi_{N+1}[1]\int \Phi_{N+1}[1]\psi_{1,x}[1]{\mathrm{d}}x=-\frac{\dot{C}(t)f_1\int
\psi_{1,x}g_1{\mathrm{d}}x }{(\int
f_{1,x}g_1{\mathrm{d}}x-C(t))^2},$$ i.e.
$$\frac{\dot{C}(t)f_1}{\int f_{1,x}g_1{\mathrm{d}}x-C(t)}\int \frac{g_1}{\int f_1g_{1,x}{\mathrm{d}}x+C(t)}(\psi_1-f_1\frac{\int \psi_{1,x}g_1{\mathrm{d}}x}{\int f_{1,x}g_1{\mathrm{d}}x-C(t)})_x{\mathrm{d}}x=-\frac{\dot{C}(t)f_1\int
\psi_{1,x}g_1{\mathrm{d}}x }{(\int
f_{1,x}g_1{\mathrm{d}}x-C(t))^2},$$ i.e., to prove
\begin{equation}
\label{314} \int \frac{g_1}{\int
f_1g_{1,x}{\mathrm{d}}x+C(t)}(\psi_1-f_1\frac{\int
\psi_{1,x}g_1{\mathrm{d}}x}{\int
f_{1,x}g_1{\mathrm{d}}x-C(t)})_x{\mathrm{d}}x=-\frac{\int
\psi_{1,x}g_1{\mathrm{d}}x }{\int f_{1,x}g_1{\mathrm{d}}x-C(t)},
\end{equation}
\begin{equation}
\label{315}
\begin{array}{ll}
&\text{the l.h.s of (\ref{314})}\\
=&\int \frac{g_1}{\int
f_1g_{1,x}{\mathrm{d}}x+C(t)}(\psi_{1,x}-\frac{f_{1,x}\int
g_1\psi_{1,x}{\mathrm{d}}x+f_1g_1\psi_{1,x}}{\int
f_{1,x}g_1{\mathrm{d}}x-C(t)}+f_1g_1f_{1,x}\frac{\int
g_1\psi_{1,x}{\mathrm{d}}x}{(\int
f_{1,x}g_1{\mathrm{d}}x-C(t))^2}){\mathrm{d}}x\\
=&\int [-\frac{g_1\psi_{1,x}}{\int f_{1,x}g_1{\mathrm{d}}x-C(t)}+\frac{g_1f_{1,x}\int g_1\psi_{1,x}{\mathrm{d}}x}{(\int f_{1,x}g_1{\mathrm{d}}x-C(t))^2}]{\mathrm{d}}x\\
=&-\int (\frac{\int g_1\psi_{1,x}{\mathrm{d}}x}{\int
f_{1,x}g_1{\mathrm{d}}x-C(t)})_x{\mathrm{d}}x\\
=&-\frac{\int g_1\psi_{1,x}{\mathrm{d}}x}{\int f_{1,x}g_1{\mathrm{d}}x-C(t)}\\
=&\text{the r.h.s of (\ref{314})}.
\end{array}
\end{equation}
This completes the proof.

{\bf{Remark}}: If $C(t)$ is not a constant, i.e.
$\frac{d}{dt}C(t)\neq 0$, the DT (\ref{312}) provides a
non-auto-B\"{a}cklund transformation between two mKPESCSs
(\ref{36}) with degree $N$ and $N+1$ respectively.

\section{The n-times Repeated Generalized Darboux Transformation for the mKPESCS}
\setcounter{equation}{0} \hskip\parindent Assuming $f_1$,
$...$,$f_n$ are $n$ arbitrary solutions  of (\ref{37}) and $g_1$,
$...$,$g_n$ are n arbitrary solutions of (\ref{38}),
$C_1(t),...,C_n(t)$ are $n$ arbitrary functions in $t$, we define
the following Wronskians:
\begin{equation}
\label{40}
\begin{array}{lll}
W_1(f_1,...,f_n;g_1,...,g_n;C_1(t),...,C_n(t))&=&det(X_{n\times
n}),\\W_2(f_1,...,f_n;g_1,...,g_n;C_1(t),...,C_n(t))&=&det(\tilde{X}_{n\times
n}),\\W_3(f_1,...,f_n;g_1,...,g_{n-1};C_1(t),...,C_{n-1}(t))&=&det(Y_{n\times
n}),\\W_4(f_1,...,f_{n-1};g_1,...,g_n;C_1(t),...,C_{n-1}(t))&=&det(\tilde{Y}_{n\times
n}),
\end{array}
\end{equation}
where
\begin{subequations}
\label{41}
\begin{equation}
\label{41.1}
     X_{i,j}=-\delta_{i,j}C_i(t)+\int g_jf_{i,x}{\mathrm{d}}x,
\end{equation}
\begin{equation}
\label{41.2} \tilde{X}_{i,j}=\delta_{i,j}C_i(t)+\int
g_{j,x}f_i{\mathrm{d}}x,\ \ \  i,j=1,...,n,
\end{equation}
\begin{equation}
\label{41.3} Y_{i,j}=-\delta_{i,j}C_i(t)+\int
g_if_{j,x}{\mathrm{d}}x,\ i=1,...,n-1,\ j=1,...,n;\ \
Y_{n,j}=f_j,\ j=1,...,n.
\end{equation}
\begin{equation}
\label{41.4} \tilde{Y}_{i,j}=\delta_{i,j}C_i(t)+\int
g_{j,x}f_i{\mathrm{d}}x,\ i=1,...,n-1,\ j=1,...,n;\ \
\tilde{Y}_{n,j}=g_j,\ j=1,...,n.
\end{equation}
\end{subequations}

\begin{lemma} Assume $f_1,...,f_n$ are solutions of (\ref{37}) and $g_1,...,g_n$ are solutions
of (\ref{38}), then for $2\leq m\leq n$, $1\leq k\leq n-m$, we
have
\begin{subequations}
\label{42}
\begin{equation}
\label{42.1}
\begin{array}{ll}
    &W_1(f_m[m-1],...,f_{m+k}[m-1];g_m[m-1],...,g_{m+k}[m-1];C_m(t),...,C_{m+k}(t))\\
    =&\frac{W_1(f_{m-1}[m-2],...,f_{m+k}[m-2];g_{m-1}[m-2],...,g_{m+k}[m-2];C_{m-1}(t),...,C_{m+k}(t))}{-C_{m-1}(t)+\int f_{m-1}[m-2]_xg_{m-1}[m-2]{\mathrm{d}}x},
\end{array}
\end{equation}

\begin{equation}
\label{42.2}
\begin{array}{ll}
&W_2(f_m[m-1],...,f_{m+k}[m-1];g_m[m-1],...,g_{m+k}[m-1];C_m(t),...,C_{m+k}(t))\\
    =&\frac{W_2(f_{m-1}[m-2],...,f_{m+k}[m-2];g_{m-1}[m-2],...,g_{m+k}[m-2];C_{m-1}(t),...,C_{m+k}(t))}{C_{m-1}(t)+\int f_{m-1}[m-2]g_{m-1}[m-2]_x{\mathrm{d}}x},
\end{array}
\end{equation}
\begin{equation}
\label{42.3}
\begin{array}{ll}
&W_3(f_m[m-1],...,f_{m+k}[m-1];g_m[m-1],...,g_{m+k-1}[m-1];C_m(t),...,C_{m+k-1}(t))\\
    =&\frac{W_3(f_{m-1}[m-2],...,f_{m+k}[m-2];g_{m-1}[m-2],...,g_{m+k-1}[m-2];C_{m-1}(t),...,C_{m+k-1}(t))}{-C_{m-1}(t)+\int f_{m-1}[m-2]_xg_{m-1}[m-2]{\mathrm{d}}x},
\end{array}
\end{equation}
\begin{equation}
\label{42.4}
\begin{array}{ll}
&W_4(f_m[m-1],...,f_{m+k-1}[m-1];g_m[m-1],...,g_{m+k}[m-1];C_m(t),...,C_{m+k-1}(t))\\
   =&\frac{W_4(f_{m-1}[m-2],...,f_{m+k-1}[m-2];g_{m-1}[m-2],...,g_{m+k}[m-2];C_{m-1}(t),...,C_{m+k-1}(t))}{C_{m-1}(t)+\int
   f_{m-1}[m-2]g_{m-1}[m-2]_x{\mathrm{d}}x}.
\end{array}
\end{equation}
\end{subequations}
\end{lemma}
This lemma can be proved in the same way as we did in
\cite{XiaoTing2004}. Then we have
\begin{theorem}
Assume that $u, \Psi_1,\cdots,\Psi_N,\Phi_1,\cdots,\Phi_N$ is a
solution of the mKPESCS (\ref{36}), $f_1,\cdots, f_n$ and
$g_1,\cdots, g_n$ are solutions of (\ref{37}) and (\ref{38})
respectively, $C_1(t),...,C_n(t)$ are $n$ arbitrary functions in
$t$. Then the $n$-times repeated generalized Darboux
transformation for (\ref{37}) is given by
\begin{subequations}
\label{48}
\begin{equation}
\label{48.1} \psi_1[n]=\frac {W_3(f_1,..., f_n, \psi_1;g_1,...,
g_n;C_1(t),...,C_n(t))}{W_1(f_1,..., f_n;g_1,...,
g_n;C_1(t),...,C_n(t))},
\end{equation}
\begin{equation}
\label{48.2} u[n]=u+\partial_x{\mathrm{ln}} \frac{W_2(f_1,...,
f_n;g_1,..., g_n;C_1(t),...,C_n(t))}{W_1(f_1,..., f_n;g_1,...,
g_n;C_1(t),...,C_n(t))},
\end{equation}
\begin{equation}
\label{48.3}
    \Psi_i[n]=\frac {W_3(f_1,..., f_n, \Psi_i;g_1,...,
g_n;C_1(t),...,C_n(t))}{W_1(f_1,..., f_n;g_1,...,
g_n;C_1(t),...,C_n(t))},
\end{equation}
\begin{equation}
\label{48.4}
    \Phi_i[n]=\frac {W_4(f_1,..., f_n;g_1,...,
g_n, \Phi_i;C_1(t),...,C_n(t))}{W_2(f_1,..., f_n;g_1,...,
g_n;C_1(t),...,C_n(t))},
\end{equation}
\begin{equation}
\label{48.5}
\begin{array}{l}
\Psi_{N+j}[n]\\
    =-\frac{1}{2}\sqrt{\dot{C}_j(t)}\frac {W_3(f_1,...,f_{j-1},f_{j+1},...,
    f_n,f_j;g_1,...,g_{j-1},g_{j+1},...,
g_n;C_1(t),...,C_{j-1}(t),C_{j+1}(t),...,C_n(t))}{W_1(f_1,...,
f_n;g_1,..., g_n;C_1(t),...,C_n(t))},
 \end{array}
\end{equation}
\begin{equation}
\label{48.6}
\begin{array}{l}
    \Phi_{N+j}[n] \\
    =\frac{1}{2}\sqrt{\dot{C}_j(t)}\frac {W_4(f_1,...,f_{j-1},f_{j+1},...,
    f_n;g_1,...,g_{j-1},g_{j+1},...,
g_n,g_j;C_1(t),...,C_{j-1}(t),C_{j+1}(t),...,C_n(t))}{W_2(f_1,...,
f_n;g_1,..., g_n;C_1(t),...,C_n(t))},
\\i=1,...,N, \quad j=1,\cdots,n.
 \end{array}
\end{equation}
\end{subequations}
Namely,
\begin{subequations}
\label{49}
\begin{equation}
\label{49.1}
    \alpha\Psi_l[n]_y=\Psi_l[n]_{xx}-2u[n]\Psi_l[n]_x,
\end{equation}
\begin{equation}
\label{49.2}
    \alpha\Phi_l[n]_y=-\Phi_l[n]_{xx}-2u[n]\Phi_l[n]_x,\
    \ l=1,...,N+n,
\end{equation}
\begin{equation}
\label{49.3}
    \alpha\psi_1[n]_y=\psi_1[n]_{xx}-2u[n]\psi_1[n],
\end{equation}
\begin{equation}
\label{49.4}
    \psi_1[n]_t=A_1(u[n])\psi_1[n]+T^1_{N+n}(\Psi[n],\Phi[n])\psi_1[n].
\end{equation}
\end{subequations}
So $u[n]$,\ $\Psi_j[n],\ \Phi_j[n],\ j=1,...,N+n$ satisfy the
mKPESCS (\ref{36}) with degree $(N+n)$.
\end{theorem}

{\bf{Proof}}: By (\ref{312}) and (\ref{42}), we have
\begin{equation}
\label{410}
\begin{array}{lll}
    \psi_1[n]&=&\frac{W_3(f_{n}[n-1],\psi_1[n-1];g_n[n-1];C_n(t))}{W_1(f_n[n-1];g_n[n-1];C_n(t))}
    \nonumber\\
    &=&\frac{W_3(f_{n-1}[n-2],f_n[n-2],\psi_1[n-2];g_{n-1}[n-2],g_n[n-2];C_{n-1}(t),C_n(t))}{-C_{n-1}(t)+\int
    f_{n-1}[n-2]_xg_{n-1}[n-2]{\mathrm{d}}x}\\
    \nonumber\\
    &&\times\frac{-C_{n-1}(t)+\int
    f_{n-1}[n-2]_xg_{n-1}[n-2]{\mathrm{d}}x}{W_1(f_{n-1}[n-2],f_n[n-2];g_{n-1}[n-2],g_n[n-2];C_{n-1}(t),C_n(t))}\nonumber\\
    &=&\cdots
    \nonumber\\
    &=&\frac {W_3(f_1,f_2,..., f_n,
\psi_1;g_1,..., g_n;C_1(t),...,C_n(t))}{W_1(f_1,..., f_n;g_1,...,
g_n;C_1(t),...,C_n(t))}.
 \end{array}
\end{equation}
Similarly we can prove (\ref{48.3}) and (\ref{48.4}) hold.

\begin{equation}
\label{411}
\begin{array}{lll}
    u[n]&=&u[n-1]+\partial_x{\mathrm{ln}}\frac{W_2(f_n[n-1];g_n[n-1];C_n(t))}{
    W_1(f_n[n-1];g_n[n-1];C_n(t))}
    \nonumber\\
    &=&u[n-2]+\partial_x{\mathrm{ln}}\frac{\int f_{n-1}[n-2]g_{n-1}[n-2]_x{\mathrm{d}}x+C_{n-1}(t)}{\int f_{n-1}[n-2]_xg_{n-1}[n-2]{\mathrm{d}}x-C_{n-1}(t)}+\partial_x{\mathrm{ln}}\frac{W_2(f_n[n-1];g_n[n-1];C_n(t))}{
    W_1(f_n[n-1];g_n[n-1];C_n(t))}
    \nonumber\\
    &=&u[n-2]+\partial_x{\mathrm{ln}}
    \frac{W_2(f_{n-1}[n-2],f_n[n-2];g_{n-1}[n-2],g_n[n-2];C_{n-1}(t),C_n(t))}{W_1(f_{n-1}[n-2],f_n[n-2];g_{n-1}[n-2],g_n[n-2];
    C_{n-1}(t),C_n(t))}
    \nonumber\\
    &=&\cdots\\
    &=&u+\partial_x{\mathrm{ln}} \frac{W_2(f_1,...,
f_n;g_1,..., g_n;C_1(t),...,C_n(t))}{W_1(f_1,..., f_n;g_1,...,
g_n;C_1(t),...,C_n(t))}.
\end{array}
\end{equation}

$$f_j[j]=f_j[j-1]-\frac{f_j[j-1]\int g_j[j-1]f_j[j-1]_x{\mathrm{d}}x}{\int g_j[j-1]f_j[j-1]_x{\mathrm{d}}x-C_j(t)}=-\frac{C_j(t)f_j[j-1]}{\int g_j[j-1]f_j[j-1]_x{\mathrm{d}}x-C_j(t)},$$
$$\Psi_{N+j}[j]=-\frac{1}{2}\frac{\sqrt{\dot{C}_j(t)}f_j[j-1]}{\int f_j[j-1]_xg_j[j-1]{\mathrm{d}}x-C_j(t)},$$
so
$$\Psi_{N+j}[j]=\frac{1}{2}\frac{\sqrt{\dot{C}_j(t)}}{C_j(t)}f_j[j].$$
So
\begin{equation}
\label{412}
\begin{array}{ll}
    &\Psi_{N+j}[n]\\
    =&\frac{W_3(f_n[n-1],\Psi_{N+j}[n-1];g_n[n-1];C_n(t))}{W_1(f_n[n-1];g_n[n-1];C_n(t))}
    \nonumber\\
    =&\cdots
    \nonumber\\
    =&\frac{W_3 (f_{j+1}[j],\cdots,f_n[j],\Psi_{N+j}[j];g_{j+1}[j],\cdots,g_n[j];C_{j+1}(t),\cdots,C_n(t))}{W_1(f_{j+1}[j],\cdots,f_n[j];g_{j+1}[j],\cdots,g_n[j];C_{j+1}(t),\cdots,C_n(t))}
    \nonumber\\
    =&\frac{\sqrt{\dot{C}_j(t)}}{2C_j(t)}\frac{W_3 (f_{j+1}[j],\cdots,f_n[j],f_j[j];g_{j+1}[j],\cdots,g_n[j];C_{j+1}(t),\cdots,C_n(t))}{W_1(f_{j+1}[j],\cdots,f_n[j];g_{j+1}[j],\cdots,g_n[j];C_{j+1}(t),\cdots,C_n(t))}
    \nonumber\\
    =&\cdots
    \nonumber\\
    =&\frac{\sqrt{\dot{C}_j(t)}}{2C_j(t)}\frac{W_3(f_1,\cdots,f_n,f_j;g_1,\cdots,g_n;C_1(t),\cdots,C_n(t))}{W_1(f_1,\cdots,f_n;g_1,\cdots,g_n;C_1(t),\cdots,C_n(t))}
    \\
    =&-\frac{1}{2}\sqrt{\dot{C}_j(t)}\frac {W_3(f_1,...,f_{j-1},f_{j+1},...,
    f_n,f_j;g_1,...,g_{j-1},g_{j+1},...,
g_n;C_1(t),...,C_{j-1}(t),C_{j+1}(t),...,C_n(t))}{W_1(f_1,...,
f_n;g_1,..., g_n;C_1(t),...,C_n(t))}.
\end{array}
\end{equation}
Similarly we can prove (\ref{48.6})
holds.\\
This completes the proof.

{\bf{Remark}}: If $C_j(t), j=1,...,n$ are not constants, i.e.
$\frac{d}{dt}C_j(t)\neq 0$, the DT (\ref{48}) provides a
non-auto-B\"{a}cklund transformation between two mKPESCSs
(\ref{36}) with degree $N$ and $N+n$ respectively.
\section{Some examples of solutions for the mKPESCS}
\setcounter{equation}{0} \hskip\parindent \ \ 1. {\bf Rational
solution. }

{\bf{Example 1}}: Rational solution with singularities
for the mKPIIESCS ($\alpha=1$).\\
If we set $\alpha=1$ in equation (\ref{36}), we get the mKPIIESCS
\begin{subequations}
\label{51}
\begin{equation}
\label{51.1}
    u_t+u_{xxx}+3D^{-1}(u_{yy})-6D^{-1}(u_y)u_x-6u^2u_x+4\sum_{i=1}^{N}(\Psi_i\Phi_i)_x=0,
\end{equation}
\begin{equation}
\label{51.2}
    \Psi_{i,y}=\Psi_{i,xx}-2u\Psi_{i,x},
\end{equation}
\begin{equation}
\label{51.3}
    \Phi_{i,y}=-\Phi_{i,xx}-2u\Phi_{i,x}.
\end{equation}
\end{subequations}
We take $u=0$, $\Phi_1=ae^{kx-k^2y}$, $\Psi_1=be^{-kx+k^2y}$,
$k,a,b\in \mathbb{R}$ as the initial solution of (\ref{51}) with
$N=1$ and let
$$f_1=(2x+8ky-96k^2t-\frac{8abt}{9k})e^{2kx+4k^2y-32k^3t-\frac{8ab}{3}t},\ \ g_1=e^{2kx-4k^2y-32k^3t+8abt},\ \ C(t)=0,$$
then by DT (\ref{312}), we get the rational solution with
singularities for the mKPIIESCS (\ref{51}) with $N=1$ as follows
\begin{subequations}
\label{52}
\begin{equation}
\label{52.1}  u[1]=\partial_x{\mathrm{ln}}\frac{\int
f_1g_{1,x}{\mathrm{d}}x}{\int g_1f_{1,x}{\mathrm{d}}x} =
\frac{8k}{(2kA+1)(2kA-1)},
\end{equation}

\begin{equation}
\label{52.2}
    \Psi_1[1]=3be^{-kx+k^2y}\frac{A+\frac{1}{6k}}{A+\frac{1}{2k}},
\end{equation}
\begin{equation}
\label{52.3}
    \Phi_1[1]=\frac{1}{3}ae^{kx-k^2y}\frac{A-\frac{1}{6k}}{A-\frac{1}{2k}},
\end{equation}
\end{subequations}
where $A=2x+8ky-96k^2t-\frac{8abt}{9k}$.\\
More generally, if we take
\begin{subequations}
\label{53}
\begin{equation}
\label{53.1}
f_i=(x+2k_iy-12k_i^2t+\frac{4abt}{(k_i+k)^2})e^{k_ix+k_i^2y-4k_i^3-\frac{4abt}{k_i+k}},
\end{equation}

\begin{equation}
\label{53.2}
   g_i=e^{k_ix-k_i^2y-4k_i^3+\frac{4k_iabt}{k_i-k}},
\end{equation}
\begin{equation}
\label{53.3} C_i(t)=0,\ \ k_i\neq \pm k, i=1,...,n, k_i+k_j\neq
0,\ \forall i,j,
\end{equation}
\end{subequations}
then (\ref{48.2}), (\ref{48.3}) and (\ref{48.4}) will give the
rational solution with multi-singularities for the mKPIIESCS with
$N=1$.

{\bf{Example 2}}: Lump solution for the mKPIESCS ($\alpha=i$).\\
If we set $\alpha=i$ in equation (\ref{36}), we get the mKPIESCS
\begin{subequations}
\label{54}
\begin{equation}
\label{54.1}
    u_t+u_{xxx}-3D^{-1}(u_{yy})-6iD^{-1}(u_y)u_x-6u^2u_x+4\sum_{i=1}^{N}(\Psi_i\Phi_i)_x=0,
\end{equation}
\begin{equation}
\label{54.2}
    i\Psi_{i,y}=\Psi_{i,xx}-2u\Psi_{i,x},
\end{equation}
\begin{equation}
\label{54.3}
    i\Phi_{i,y}=-\Phi_{i,xx}-2u\Phi_{i,x}.
\end{equation}
\end{subequations}
We take $u=0$, $\Phi_1=ae^{-ikx-ik^2y}$, $\Psi_1=be^{ikx+ik^2y}$,
$k,a,b\in \mathbb{R}$ as the initial solution of (\ref{54}) with
$N=1$ and let
$$f_1=(x-2ly+12l^2t-\frac{4abkti}{(k+l)^2})e^{-ilx+il^2y-4il^3t-\frac{4ablt}{k+l}},\ \ g_1=e^{-ilx-il^2y-4il^3t+\frac{4ablt}{l-k}},$$
$l\in \mathbb{R}$, $l\neq \pm k$ and \ \ $C(t)=0,$ then by DT
(\ref{312}), we get the 1-lump solution for the mKPIESCS
(\ref{54}) with $N=1$ as follows
\begin{subequations}
\label{55}
\begin{equation}
\label{55.1} u[1]=\partial_x{\mathrm{ln}}\frac{\int
f_1g_{1,x}{\mathrm{d}}x}{\int g_1f_{1,x}{\mathrm{d}}x}=
\frac{4li}{1+(2lA)^2},
\end{equation}

\begin{equation}
\label{55.2}
    \Psi_1[1]=be^{kxi+k^2yi}\frac{-2Al(k+l)+i(k-l)}{2lA(k-l)+i(k-l)},
\end{equation}
\begin{equation}
\label{55.3}
    \Phi_1[1]=ae^{-kxi-k^2yi}\frac{2lA(k^2-l^2)+(k-l)^2i}{-2lA(l+k)^2+i(k+l)^2},
\end{equation}
\end{subequations}
where $A=x-2ly+12l^2t-4ab\frac{ikt}{(k+l)^2}$.\\
More generally, if we take
\begin{subequations}
\label{56}
\begin{equation}
\label{56.1}
f_j=(x-2l_jy+12l_j^2t+\frac{4l_jabt}{(l_j+k)^2})e^{-il_jx+il_j^2y-4il_j^3t-\frac{4il_jabt}{l_j+k}},
\end{equation}

\begin{equation}
\label{56.2}
   g_j=e^{-il_jx-il_j^2y-4il_j^3t+\frac{4il_jabt}{l_j-k}},
\end{equation}
\begin{equation}
\label{56.3} C_j(t)=0,\ \ l_j\neq \pm k, j=1,...,n, l_m+l_j\neq
0,\ \forall m,j,
\end{equation}
\end{subequations}
then (\ref{48.2}), (\ref{48.3}) and (\ref{48.4}) will give the
multi-lump solution for the mKPIESCS with $N=1$.\\
2. {\bf Soliton solution.}

{\bf{Example 3}}: Soliton solution for
the mKPIIESCS.\\
We take $u=0$ as the initial solution for the mKPIIESCS (\ref{51})
with $N=0$ and let
$$f_1=e^{kx+k^2y-4k^3t}=e^{\xi_1}, \ \ \ g_1=e^{lx-l^2y-4l^3t}=e^{\xi_2},\ \ \ C(t)=e^{2\beta(t)},$$
where$k,l\in \mathbb{R}$, $k+l\neq 0$, and $\beta(t)$ is an
arbitrary function in $t$. Then by DT (\ref{312}), we get the
1-soliton solution for the mKPIIESCS (\ref{51}) with $N=1$ as
follows
\begin{subequations}
\label{57}
\begin{equation}
\label{57.1} u[1]=\partial_x{\mathrm{ln}}\frac{\int
f_1g_{1,x}{\mathrm{d}}x+C(t)}{\int g_1f_{1,x}{\mathrm{d}}x-C(t)}=
-\frac{k+l}{(\frac{l}{k+l}e^{\eta}-e^{-\eta})(\frac{k}{k+l}e^{\eta}+e^{-\eta})},\
\ \eta=\frac{\xi_1+\xi_2}{2}-\beta(t),\\
\end{equation}

\begin{equation}
\label{57.2}
    \Psi_1[1]=-\frac{1}{2}f_1\frac{\sqrt{\dot{C}(t)}}{\int
    f_{1,x}g_1{\mathrm{d}}x-C(t)}=-\frac{\sqrt{2\dot{\beta}(t)}}{2}\frac{e^{\xi_1+\beta(t)}}{\frac{k}{k+l}e^{\xi_1+\xi_2}-e^{2\beta(t)}},
\end{equation}
\begin{equation}
\label{57.3}
    \Phi_1[1]=\frac{1}{2}g_1\frac{\sqrt{\dot{C}(t)}}{\int
    g_{1,x}f_1{\mathrm{d}}x+C(t)}=\frac{\sqrt{2\dot{\beta}(t)}}{2}\frac{e^{\xi_2+\beta(t)}}{\frac{l}{k+l}e^{\xi_1+\xi_2}+e^{2\beta(t)}}.
\end{equation}
\end{subequations}
More generally, if we take
\begin{equation}
\label{58} f_i=e^{k_ix+k_i^2y-4k_i^3t},\ \ \
g_i=e^{l_ix-l_i^2y-4l_i^3t},\ \ \ C_i(t)=e^{\beta_i(t)},i=1,...,n,
\end{equation}
where $k_i,l_i\in \mathbb{R}$,\ $ k_i+l_j\neq 0,\ \forall i,j$ ,
then (\ref{48.2}), (\ref{48.5}) and (\ref{48.6}) will give the
n-soliton solution for the mKPIIESCS with $N=n$.

{\bf{Example 4}}: Soliton solution for
the mKPIESCS.\\
We take $u=0$ as the initial solution for the mKPIESCS (\ref{54})
with $N=0$ and let
$$f_1=e^{-ikx+ik^2y-4ik^3t}, \ \ \ g_1=e^{i\bar{k}x-i\bar{k}^2y+4i\bar{k}^3t},\ \ \ C(t)=ie^{2\beta(t)}$$
where $k\in \mathbb{C}$ and $\beta(t)$ is an arbitrary function in
$t$.\\
Set \ \ $k=\mu-i\nu$,\ \ \ \ $\mu,\nu\in\mathbb{R}$,\ \ \ \
$\nu\neq
 0,$\\
then
$$f_1=e^{\theta+\eta},\ \ \ \ g_1=\bar{f}_1=e^{-\theta+\eta}$$
where
$$\theta=-i\mu x+i(\mu^2-\nu^2)y-4i(\mu^3-3\mu\nu^2)t,\ \ \ \eta=-\nu x+2\mu\nu y+4\nu(\nu^2-3\mu^2)t.$$
Then by DT (\ref{312}), we get the 1-soliton solution for the
mKPIESCS (\ref{54}) with $N=1$ as follows
\begin{subequations}
\label{59}
\begin{equation}
\label{59.1} u[1]=\partial_x{\mathrm{ln}}\frac{\int
f_1g_{1,x}{\mathrm{d}}x+C(t)}{\int g_1f_{1,x}{\mathrm{d}}x-C(t)}=
\frac{2\nu i}{\frac{1}{4}e^{2f}+(e^{-f}-\frac{\mu}{2\nu}e^f)^2},\ \ f=\eta-\beta(t)\\
\end{equation}

\begin{equation}
\label{59.2}
    \Psi_1[1]=-\frac{1}{2}f_1\frac{\sqrt{\dot{C}(t)}}{\int
    f_{1,x}g_1{\mathrm{d}}x-C(t)}=\frac{\sqrt{\dot{\beta}(t)}(1-i)e^{\theta+\nu x+2\mu\nu y+4\nu^3t+12\mu^2\nu t}}{(i\nu-\mu)e^{8\nu^3t+4\mu\nu y}+2\nu e^{\beta(t)+24\mu^2\nu t+2\nu x}},
\end{equation}
\begin{equation}
\label{59.3}
    \Phi_1[1]=\frac{1}{2}g_1\frac{\sqrt{\dot{C}(t)}}{\int
    g_{1,x}f_1{\mathrm{d}}x+C(t)}=\frac{\sqrt{\dot{\beta}(t)}(i-1)e^{-\theta+\nu x+2\mu\nu y+4\nu^3t+12\mu^2\nu t}}{(i\nu+\mu)e^{8\nu^3t+4\mu\nu y}-2\nu e^{\beta(t)+24\mu^2\nu t+2\nu x}},
\end{equation}
\end{subequations}
More generally, if we take
\begin{equation}
\label{510} f_j=e^{-ik_jx+ik_j^2y-4ik_j^3t},\ \ \
g_j=e^{i\bar{k}_jx-i\bar{k}_j^2y+4i\bar{k}_j^3t},\ \ \
C_j(t)=ie^{2\beta_j(t)},j=1,...,n,
\end{equation}
where $k_j=\mu_j+i\nu_j$, $\mu_j,\nu_j \in\mathbb{R}$, $k_j\neq
\bar{k}_m, \ \forall j,m$ and $\beta_j(t)$, $j=1,...,n,$ are
arbitrary functions in $t$, then (\ref{48.2}), (\ref{48.5}) and
(\ref{48.6}) will give the n-soliton solution for the mKPIESCS
with $N=n$.

3. {\bf Solutions of breather type.}

{\bf{Example 5}}: Solutions
of breather type for the mKPIESCS.\\
We take $u=0$ as the initial solution for the mKPIESCS (\ref{54})
with $N=0$. If we take
$$f_j=e^{-i\lambda_jx+i\lambda_j^2y-4i\lambda_j^3t}, \ \ \ g_j=e^{i\xi_jx-i\xi_j^2y+4i\xi_j^3t},\ \ \ C_j(t)=ie^{2\beta_j(t)},\ \ j=1,...,2n$$
where
$$(\lambda_1,...,\lambda_{2n})=(k_1,...,k_n;l_1,...,l_n),\ \ \ \ (\xi_1,...,\xi_{2n})=(\bar{l}_1,...,\bar{l}_n;\bar{k}_1,...,\bar{k}_n),$$
$k_j,l_j\in \mathbb{C}$, $Im(k_j)\neq 0$, $Im(l_j)\neq 0$,
$l_m\neq \bar{k}_j$, $\forall m,j$,\\
we will get the solutions of breather type for the mKPIESCS by
(\ref{48.2}), (\ref{48.5}) and
(\ref{48.6}).\\
For example, if we choose $n=1$ and
$$\lambda_1=k_1=-bi,\ \ \lambda_2=l_1=-di,\ \ \ \xi_1=\bar{\lambda}_2=di,\ \ \ \xi_2=\bar{\lambda}_1=bi,\ \ \ C_1(t)=C_2(t)=ie^{2t},$$
we will get the following solution of mKPIESCS
\begin{subequations}
\label{511}
\begin{equation}
\label{511.1}
u[2]
=-\frac{8(b+d)^2\mathrm{cos}\theta e^fi}{4(b+d)^2e^{2t}+4(b^2-d^2)e^f+(b-d)^2e^{2f-2t}},\\
\end{equation}

\begin{equation}
\begin{array}{lll}
\label{511.2}
    \Psi_1[2]
    &=&\frac{1+i}{2}e^{\eta_1+\theta_2i+t}\frac{\frac{d-b}{2(d+b)}e^f+e^{2t}\mathrm{sin}\theta-ie^{2t}\mathrm{cos}\theta}{e^{4t}-e^{f+2t}\mathrm{sin}\theta\frac{b-d}{b+d}+\frac{(b-d)^2}{4(b+d)^2}e^{2t}+e^{f+2t}\mathrm{cos}\theta
i},
\end{array}
\end{equation}

\begin{equation}
\begin{array}{lll}
\label{511.3}

    \Psi_2[2]
    &=&\frac{1+i}{2}e^{\eta_2+\theta_1i+t}\frac{\frac{b-d}{2(d+b)}e^f-e^{2t}\mathrm{sin}\theta-ie^{2t}\mathrm{cos}\theta}{e^{4t}-e^{f+2t}\mathrm{sin}\theta\frac{b-d}{b+d}+\frac{(b-d)^2}{4(b+d)^2}e^{2t}+e^{f+2t}\mathrm{cos}\theta
i},
\end{array}
\end{equation}

\begin{equation}
\begin{array}{lll}
\label{511.4}

    \Phi_1[2]
    &=&-\frac{1+i}{2}e^{\eta_2-\theta_1i+t}\frac{\frac{b+d}{2(b-d)}e^f-e^{2t}\mathrm{sin}\theta+ie^{2t}\mathrm{cos}\theta}{e^{4t}-e^{f+2t}\mathrm{sin}\theta\frac{b-d}{b+d}+\frac{(b-d)^2}{4(b+d)^2}e^{2t}-e^{f+2t}\mathrm{cos}\theta
i},
\end{array}
\end{equation}
\begin{equation}
\begin{array}{lll}
\label{511.5}

   \Phi_2[2]
    &=&-\frac{1+i}{2}e^{\eta_1-\theta_2i+t}\frac{\frac{b+d}{2(d-b)}e^f+e^{2t}\mathrm{sin}\theta+ie^{2t}\mathrm{cos}\theta}{e^{4t}-e^{f+2t}\mathrm{sin}\theta\frac{b-d}{b+d}+\frac{(b-d)^2}{4(b+d)^2}e^{2t}-e^{f+2t}\mathrm{cos}\theta
i},
\end{array}
\end{equation}
\end{subequations}
where
$$f=-(b+d)x+4(b^3+d^3)t,\ \ \ \theta=(d^2-b^2)y,\ \ \ \eta_1=-bx+4b^3t,\ \ \ \eta_2=-dx+4d^3t,$$
$$\theta_1=-b^2y,\ \ \ \theta_2=-d^2y.$$
$u[2]$ is periodic in $y$ and has soliton
behavior along the coordinate $x$.\\
Similarly, we can get the solution of breather type for the
mKPIIESCS.

{\bf{Remark}}: In {\bf{Example 1}} and {\bf{Example 2}}, when
$a=b=0$, the solutions obtained above will degenerate to the
solutions of the corresponding mKP equations respectively. In
{\bf{Example 3}}, {\bf{Example 4}} and {\bf{Example 5}}, when
$C(t)(C_j(t))$ are taken to be constant(s), i.e.
$\frac{d}{dt}C(t)(\frac{d}{dt}C_j(t))=0$, the solutions obtained
above will also degenerate to the solutions of the corresponding
mKP equations respectively \cite{Konopelchenko92}.


\section*{Acknowledgment}\hskip\parindent
This work was supported by the Chinese Basic Research Project
"Nonlinear Science".

\hskip\parindent

\end{document}